\documentclass{elsart}

\usepackage{graphicx}

\begin{document}

\begin{frontmatter}

\title{On a perturbation treatment of a model for MHD viscous flow}

\author{Francisco M. Fern\'{a}ndez\thanksref{FMF}}

\address{INIFTA (UNLP, CCT La Plata-CONICET), Divisi\'{o}n Qu\'{i}mica Te\'{o}rica,\\
Blvd. 113 y 64 (S/N), Sucursal 4, Casilla de Correo 16,\\
1900 La Plata, Argentina}

\thanks[FMF]{e--mail: fernande@quimica.unlp.edu.ar}

\begin{abstract}
We discuss the solution of a nonlinear ordinary differential equation that
appears in a model for MHD viscous flow caused by a shrinking sheet. We
propose an accurate numerical solution and derive simple analytical expressions.
Our results suggest that a recent perturbation treatment of the same
problem exhibits a pathological behaviour and conjecture its
probable cause.
\end{abstract}

\end{frontmatter}

\section{Introduction}

There has recently been enormous interest in the so called homotopy
perturbation methods. In particular there is an open controversy between the
developers of homotopy perturbation method (HPM) and homotopy analysis
method (HAM)\cite{H04,L05}. One of the controversial points is related to
the appearance of some ``secular'' terms in the perturbation solutions.
Whereas He\cite{H04} proposed their removal, Liao\cite{L05}
argued that they are harmless because they vanish as the variable increases
towards infinity.

Sajit and Hayat\cite{SH09} have recently discussed the MHD viscous flow due
to a shrinking sheet. They converted the model partial differential
equations into a nonlinear ordinary differential equation which they solved
by means of HAM. The authors discussed the convergence of the perturbation
approach and obtained apparently accurate results for several values of the
model parameters. In fact, they concluded that ``The obtained HAM solution
is valid for all values of the suction parameter and Hartman number.''

The purpose of this paper is to analyze the accuracy of those HAM results
because we believe that they may cast light on the abovementioned discussion
about the ``secular'' terms. In Sec.~\ref{sec:Hankel} we consider an
accurate numerical calculation, in Sec.~\ref{sec:analytical} we derive
simple approximate analytical expressions, and in Sec.~\ref{sec:Results} we
discuss the results for a particular choice of the model parameters and draw
conclusions.

\section{Accurate numerical calculation}

\label{sec:Hankel}

Since we are not interested in the validity and usefulness of the model we
just concentrate on the equation that the authors solved approximately by
means of HAM. By means of an appropriate transformation Sajit and Hayat\cite
{SH09} converted the model partial differential equations into the ordinary
nonlinear differential equation
\begin{eqnarray}
&&f^{\prime \prime \prime }(\eta )-M^{2}f^{\prime }(\eta )-f^{\prime }(\eta
)^{2}+mf(\eta )f^{\prime \prime }(\eta )=0  \nonumber \\
&&f(0)=s,\;f^{\prime }(0)=-1  \nonumber \\
&&\lim_{\eta \rightarrow }f^{\prime }(\eta )=0  \label{eq:equation}
\end{eqnarray}
where a prime indicates differentiation with respect to the variable $\eta $
and $M$, $m$ and $s$ are model parameters. The main problem is to obtain the
value of $f^{\prime \prime }(0)$ that is consistent with the condition at
infinity. Once we have it, then we can resort to any numerical integration
routine to obtain the solution for all $\eta >0$.

In order to determine $f^{\prime \prime }(0)=\alpha $ accurately we apply
the Hankel--Pad\'{e} method developed some time ago that proved successful
for the treatment of two--point boundary value problems\cite{AF07,F08,AH09}.
It consists of expanding the solution in a Taylor series about $\eta =0$%
\begin{equation}
f(\eta )=\sum_{j=0}^{\infty }f_{j}(\alpha )\eta ^{j}  \label{eq:f_series}
\end{equation}
and then calculating the roots of the Hankel determinant $H_{D}^{d}(\alpha )$
with matrix elements $f_{i+j+d}(\alpha )$, $i,j=1,2,\ldots ,D$. The
calculation is straightforward because the matrix elements and,
consequently, the Hankel determinant are polynomial functions of the unknown
parameter $\alpha $. As shown in earlier applications of the method\cite
{AF07,F08,AH09} one expects to find a sequence of roots $\alpha _{D}$, $%
D=2,3,\ldots $, that converges towards the appropriate value of $f^{\prime
\prime }(0)$. As said above, once we have a sufficiently accurate value of
this parameter then we can apply any numerical integration algorithm and
obtain $f(\eta )$ for all $\eta >0$. Alternatively, in some cases the Pad%
\'{e} approximants, on which the method is based, give sufficiently accurate
results\cite{F08}.

\section{Approximate analytical expressions}

\label{sec:analytical}

Approximate analytical solutions to the equations of a physical model
commonly provide greater insight into the nature of the phenomenon under
investigation.
The analytical expressions provided by HAM\cite{SH09} appear to be so
complicated that one can only use them within a computer algebra system. In
this sense this kind of solution is not much different from the results
provided by the numerical integration routines that are also built in most
such software packages. The purpose of this section is to provide simple
analytical expressions for the straightforward discussion of the MHD model.

As $\eta $ increases $f^{\prime }(\eta )^{2}$ is expected to be smaller than
the other terms and therefore  $f(\eta )$ will behave approximately as the
solution of $f^{\prime \prime \prime }(\eta )-M^{2}f^{\prime }(\eta
)+mf_{\infty }f^{\prime \prime }(\eta )=0$, where $f_{\infty }=f(\eta
\rightarrow \infty )$. In other words, we expect that
$f^\prime(\eta )\approx be^{-\beta \eta }$
for sufficiently large $\eta $. Therefore, it
seems reasonable to try the ansatz
\begin{equation}
f^{[N]}(\eta )=\sum_{j=0}^{N}b_{j}e^{-\beta j\eta },\, N=1,2,\ldots
\label{eq:f_ansatz}
\end{equation}
If we substitute it into the differential equation (\ref{eq:equation}) we
obtain an expression of the form
\begin{equation}
\sum_{j=1}^{2N}R_{j}(b_{1},b_{2},\ldots ,b_{N},\beta )e^{-\beta j\eta }=0
\label{eq:R}
\end{equation}
Therefore, the optimal values of the adjustable parameters $b_{j}$ and $%
\beta $ should be solutions to
\begin{eqnarray}
&&f^{[N]}(0)=s,\;f^{[N]\prime }(0)=-1  \nonumber \\
&&R_{j}(b_{1},b_{2},\ldots ,b_{N},\beta )=0,\;j=1,2,\ldots ,N  \label{eq:eqs}
\end{eqnarray}

Since it is our purpose, for the reasons already given above, to keep the
results as simple as possible we just consider the first two approximation
orders explicitly. For $N=1$ and $N=2$ we easily obtain
\begin{equation}
b_{0}=s-\frac{1}{\beta },\;b_{1}=\frac{1}{\beta },\;\beta =\frac{\sqrt{%
4M^{2}+m^{2}s^{2}-4m}+ms)}{2}  \label{eq:b,beta_[1]}
\end{equation}
and
\begin{eqnarray}
&&b_{0}=\frac{\beta ^{2}-M^{2}}{m\beta },\;b_{1}=\frac{2(M^{2}-\beta
^{2})+m(2\beta s-1)}{m\beta },\;b_{2}=\frac{\beta ^{2}-M^{2}+m(1-\beta s)}{%
m\beta }  \nonumber \\
&&4\beta ^{4}-4\beta ^{3}ms(2-m)-2\beta
^{2}[2m(m^{2}s^{2}-ms^{2}-1)-M^{2}(3m-4)]  \nonumber \\
&&-2\beta ms[M^{2}(5m-4)-2m(m-1)]  \nonumber \\
&&-2M^{4}(3m-2)-2M^{2}m(2-3m)-m^{2}(m-1)=0  \label{eq:b,beta_[2]}
\end{eqnarray}
respectively.

It is probable that this approach, or somewhat similar to it, had been used
in the past. However, it is useful for our purposes and we are not aware
that it had been applied to the problem discussed by Sajid and Hayat\cite
{SH09}.

\section{Results and discussion}

\label{sec:Results}

Sajid and Hayat\cite{SH09} analyzed the form of $f^{\prime }(\eta )$ for
several values of the model parameters. Here we simply consider the case $%
M=m=2$ for the largest values of $s$ for which $f^{\prime }(\eta )$ appears
to exhibit a maximum according their Figure 2.

\begin{figure}[H]
\begin{center}
\includegraphics[width=9cm]{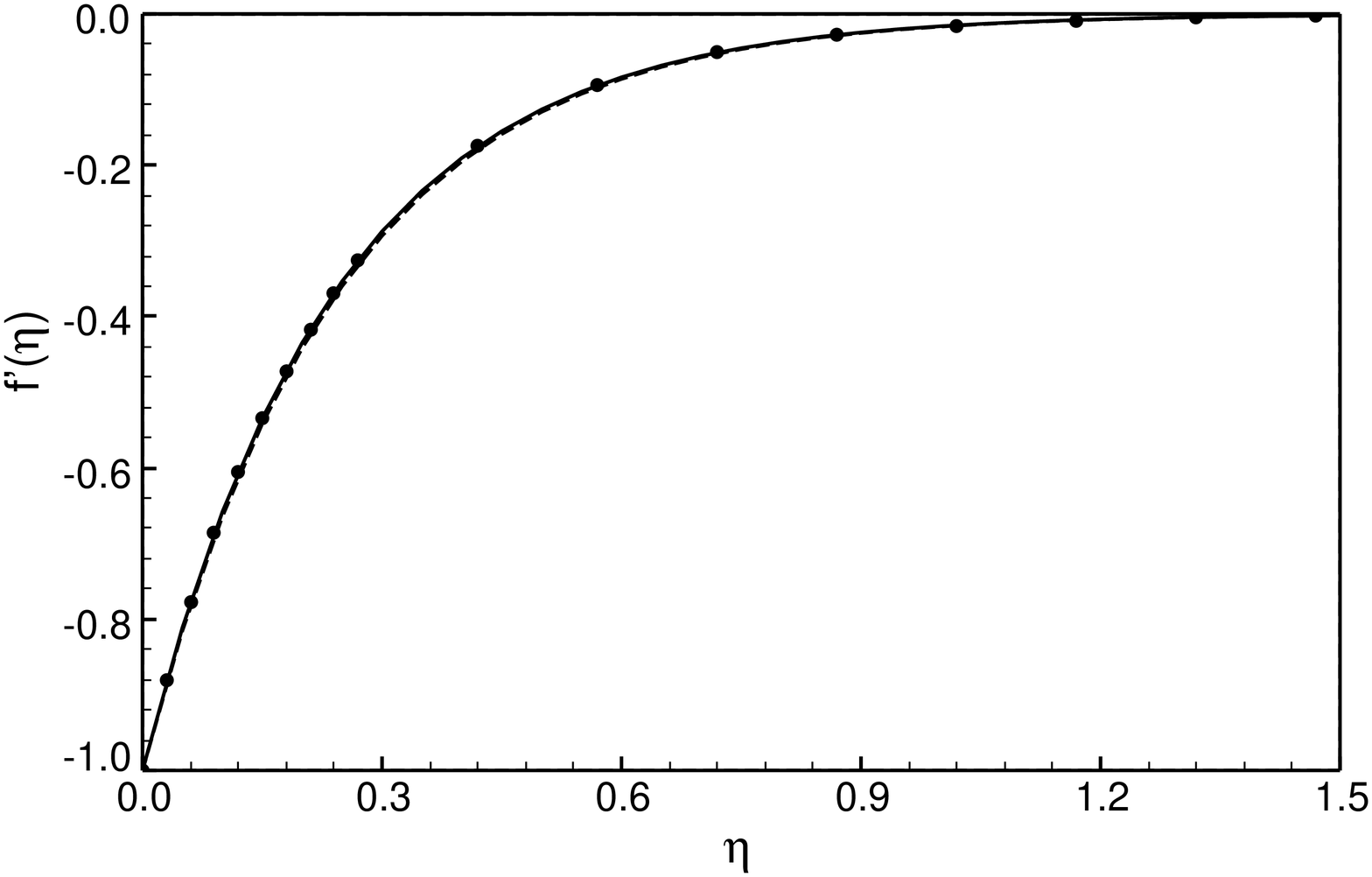}
\end{center}
\caption{Numerical results (circles), first--order (dashed line) and
second--order (solid line) analytical expressions for $f^{\prime}(\eta)$.}
\label{fig:MHD}
\end{figure}

For $s=1.8$ a sequence of roots $\alpha _{D}$ of the Hankel determinant $%
H_{D}^{d}(\alpha )$ with $d=1$ and $D=2,\ldots ,30$ suggests that $%
\alpha =4.20411340$ (the reader may find examples of the rate of convergence
of the method elsewhere\cite{AF07,F08,AH09}). Sajid and Hayat\cite{SH09}
only showed values of $f^{\prime \prime }(0)$ for $M=2$, $s=1$ and $m=1,2$;
therefore, their result for $s=1.8$ is not available for comparison. The
correct behaviour of the numerical Runge--Kutta solution for large values of
$\eta $ provides an additional confirmation of the accuracy of that value of
$f^{\prime \prime }(0)$. Fig.~\ref{fig:MHD} shows such numerical results and
also those given by the analytical expressions (\ref{eq:b,beta_[1]})
and (\ref{eq:b,beta_[2]}). Our simple analytical
expressions already provide satisfactory results for all values of $\eta $ as well as
the following acceptable estimates of $f^{\prime \prime }(0)$: $\alpha
^{[1]}=\sqrt{131}/5+9/5\approx 4.1$ and $\alpha ^{[2]}\approx 4.198$. We
appreciate that the accuracy increases with $N$ (at least for the first two
approximations). Besides, the fact that $b_{1}\approx 0.238\gg b_{2}\approx
0.00309$ for $N=2$ suggests a remarkable convergence rate.

Fig.~\ref{fig:MHD} clearly shows that neither the approximate analytical
expressions nor the accurate numerical results exhibit a maximum. Therefore,
we conclude that the maxima found by Sajid and Hayat\cite{SH09} when increasing
the suction parameter are merely
artifacts of the HAM. In our opinion, such spurious maxima are probably
caused by the ``secular'' terms of the form $\eta ^{k}e^{-n\eta }$ shown in
Eq.~(24) of Ref.~\cite{SH09} and discussed by He an Liao\cite{H04,L05}.
Although those terms certainly vanish as $\eta \rightarrow \infty $ they may
have some non--negligible undesirable effect for moderate values of $\eta $.

Summarizing: we have verified once more that the Hankel--Pad\'{e} method is
useful for the treatment of two--point boundary value problems by obtaining
an accurate value of the unknown parameter appearing in the nonlinear
equation for a MHD viscous flow model\cite{SH09}. It is necessary for a
successful application of any numerical integration routine. We have also
derived simple accurate analytical expressions that may be useful for the
discussion of the physics of the problem. Both the analytical expressions
and the numerical results have revealed a pathological behaviour of the
results produced by the much more elaborate approach called HAM\cite{SH09}.
In this way
we hope to have settled the argument about the ``secular'' terms in the HAM
expressions\cite{H04,L05}. Perhaps, the HAM users may want to verify our
conjecture and throw some more light on the subject.

\end{document}